# Reserved or On-Demand Instances? A Revenue Maximization Model for Cloud Providers

Michele Mazzucco*†, Marlon Dumas*†
*Institute of Computer Science, University of Tartu, Estonia
‡Software Technology and Applications Competence Centre (STACC), Estonia

***Abstract*—We examine the problem of managing a server farm in a way that attempts to maximize the net revenue earned by a cloud provider by renting servers to customers according to a typical Platform-as-a-Service model. The Cloud provider offers its resources to two classes of customers: 'premium' and 'basic'. Premium customers pay upfront fees to reserve servers for a specified period of time (e.g. a year). Premium customers can submit jobs for their reserved servers at any time and pay a fee for the server-hours they use. The provider is liable to pay a penalty every time a 'premium' job can not be executed due to lack of resources. On the other hand, 'basic' customers are served on a best-effort basis, and pay a server-hour fee that may be higher than the one paid by premium customers. The provider incurs energy costs when running servers. Hence, it has an incentive to turn off idle servers. The question of how to choose the number of servers to allocate to each pool (basic and premium) is answered by analyzing a suitable queuing model and maximizing a revenue function. Experimental results show that the proposed scheme adapts to different traffic conditions, penalty levels, energy costs and usage fees.**

## I. Introduction

Cloud providers who operate large and medium-scale data centers and offer their services on a pay-per-usage basis face a challenging problem of running data centers in an energy-efficient manner while ensuring that customers' expectations in terms of availability are met. Given that the amount of power consumed by an idle server is about 65% of its peak consumption [1], the only way to *significantly* reduce power consumption is to improve the server farm's utilization, either by switching off unnecessary servers, or by serving more customers with the same amount of resources. In this paper, we focus on the former strategy.

Cloud providers benefit from significant economies of scale, since they operate large infrastructures shared by users with very different workloads. In this context, workload peaks are not highly correlated, allowing the provider to safely increase system utilization [2]. Nonetheless, while cloud providers are expected to offer virtually unlimited computing capacity on-demand, in practice this is not always possible, particularly for large-scale jobs. Accordingly, there is an incentive for cloud providers to segment between customers who "reserve" servers upfront – and thus have the right to expect these instances to be available anytime – and customers who consume resources on-demand and are served on a best-effort basis. Amazon EC2[1] for example distinguishes between reserved instances (servers) and on-demand ones. Customers of reserved instances pay an upfront fee for the reservation and a per-hour usage fee . On-demand instances on the other hand only require a per-hour usage fee, which is higher than the one for reserved instances.

[1] http://aws.amazon.com/ec2/

In this paper we propose and evaluate a model for maximizing the net revenue of a cloud provider, where net revenue is defined as the fees for server usage, minus energy costs and penalties paid by the provider for service unavailability. The model considers 'premium' and 'basic' customers: the former require availability guarantees, and are entitled to receive a monetary compensation every time their job can not be executed due to a shortage in computing resources, while the latter receive a best-effort service and are not entitled to compensation if their job is refused.

The revenue maximization problem described here does not appear to have been studied before. Perhaps the most similar related work has been done in [3]. However that paper considers a single class of customers, with no penalties due to unavailability. [4] discusses a queuing model used for managing the power consumption of a service with a given Service Level Agreement (SLA). In [5], [6] customers are assumed to have a certain amount of patience, *i.e.*, if no server is available, they are willing to wait for a limited (and unknown) amount of time, while the provider is not liable to pay any penalty if a transaction is aborted due to the lack of resources. [7] proposes some allocation strategies aiming at reducing the Energy-Response time Product (ERT), but these strategies do not consider the monetary impact of lost jobs. [8] discusses the resource allocation problem in multi-tier virtualized systems with the goal of meeting the QoS requirements while minimizing energy costs. However, it does not consider the issue of paying penalties for unavailability.

Apart from the aforementioned papers, most prior studies have tried to optimize energy consumption as the primary variable, even at a cost of performance degradation. However, for a hosting center it is more important to meet the required SLA (since that is the revenue source) and reduce energy consumption when possible, and only to the extent that the reduced energy consumption is not overshadowed by penalties that need to be paid for resource unavailability or degraded performance. The main novelty of the study presented here is that it takes into account penalties due to unavailability, and it distinguishes between customers who are entitled to penalty payments due to unavailability, and those who are served on a best-effort basis.

The specific problem formulation and its associated mathematical model are presented in Section II. Section III discusses the policies for service allocation, while a number of experiments comparing the policies under different parameters and loads are reported in Section IV. Finally, we conclude the paper in Section V.

## II. The Model

At a given point in time, the provider has a cluster of $S$ identical processors/cores (hereon called the *servers*), $n$ running and $(S-n)$ switched off. The provider offers each server for a lease, and a customer who rents a server (*e.g.*, by running a virtual machine on it) is essentially creating a job. The size of the job is the length of the lease, and since the client decides when to terminate the lease, the job size is not known a priori. Servers are not shared, so each server can handle at maximum one job at any given time. As discussed in [3], given that the power drained by each CPU is a *linear function* of the load, the model we propose here can be applied to a scenario where multiple virtual machines are running on a single physical CPU. If, once a server has finished processing a request, no other jobs enter the system, the server becomes idle, *i.e.*, it consumes energy without generating any revenue.

### A. Revenue per Unit Time

The provider offers each server for a lease to two types of customers: '*premium*' (type 1), and '*basic*' (type 2). Type 2 customers pay a fee of $c_2$ per time unit (determining the amount of charge is outside the scope of this paper). If there is no server available at the instant when a type 2 job enters the system, this type 2 job is blocked and lost, without affecting future arrivals. On the other hand, 'premium' customers pay a certain upfront fee to reserve a server. This fee is irrelevant to the model. Once a server has been reserved, the customer has no further obligation (*e.g.*, he/she might decide not to use the server, and in that case no further payment is made). Like type 2 jobs, type 1 jobs are also charged in proportion to their length, at a rate of $c_1$ per unit time. One would expect that $c_2 > c_1$, but not necessarily. Importantly, if a type 1 job is rejected because no server is available, the provider is liable to pay a penalty of $d$.

Since running servers consume electricity, which costs \$ $r$ per $kWh$, the service provider tries to optimize its profits by means of a *resource allocation policy* which controls how many servers will run. The extreme values, $n = 0$ and $n = S$, correspond to switching respectively *off*, or *on*, all available servers. In order to deal with time-varying user demand the provider should be able to dynamically change the number of running servers in response to changes in user demand. Our approach is to periodically invoke a resource allocation policy that, by means of traffic estimates, determines the number of servers to run. During the intervals between consecutive policy invocations, the number of running servers remains constant. Those intervals, which will be referred to as *observation windows* or *epochs*, are used to collect traffic statistics and obtain current estimates of traffic parameters, which are used by the allocation policy at the next decision epoch.

Different metrics can be used to measure the performance of a computing system, including average or $n$-th percentile of response time, throughput, or metrics that are more specific to the area of power efficiency such as Energy-Response time Product (ERP) or Energy-Delay Product (EDP) [7]. But ultimately, as far as the service provider is concerned, the performance of the system is measured by the average revenue, $R$, earned per unit time, which in our setting can be decomposed as follows:

$$R = R_1 + R_2, \tag{1}$$

where $R_1$ is the revenue generated by type 1 jobs and $R_2$ is the profit generated by 'basic' customers.

*1) Isolated subsystems:* One possible way of structuring such a system is be to treat the two subsystems (*i.e.*, 'premium' and 'basic' customers) in isolation of each other, that is, to use $n_1$ servers for servicing type 1 jobs and $n_2$ servers for running type 2 jobs ($n_1 + n_2 = n$). If that is the case each subsystem behaves as described in [3]. Hence $R_1$ and $R_2$ can be estimated as

$$R_1 = \frac{c_1}{\mu} T_1 - rP_1 - D, \tag{2}$$

$$R_2 = \frac{c_2}{\mu} T_2 - rP_2, \tag{3}$$

where $c_i/\mu$ is the average charge paid by customers of type $i$ for running their job, $T_i$ is the throughput of subsystem $i$, $P_i$ is the average amount of electricity consumed by powered on servers of type $i$ [3], and $D$ is the average penalty paid per unit time to 'premium' customers due to the lack of resources.

Another option would be to use a single pool of servers to execute both 'premium' and 'basic' jobs. In that case, however, one would expect that the number of type 1 jobs lost under heavy load would be very high, as they would have to compete for the available resources with type 2 jobs.

*2) Overflow subsystem:* Hence, we consider a hybrid system. Under this model jobs of type 1 are offered to type 1 servers. However, if none of the $n_1$ running servers is idle the incoming jobs of type 1 are offered to the second subsystem. The 'premium' traffic directed to the second queue is called *overflow traffic*. It is worth stressing that 'premium' jobs are routed to the second queue only when the primary queue is busy, see Figure 1.

In case the electricity cost changes over the time (*e.g.*, see [9]) each policy invocation should use a different value of $r$.

### B. Throughput Estimation

In order to estimate parameters $T_i$ and $D$ in Equations (2) and (3), it is necessary to have a quantitative model of user demand and service provision. Assume that jobs of type $i$ enter the system according to and independent Poisson process with rate $\lambda_i$, and that the service times are distributed with rate $1/\mu$. Hence, the offered load of type $i$ would be $\rho_i = \lambda_i/\mu$.

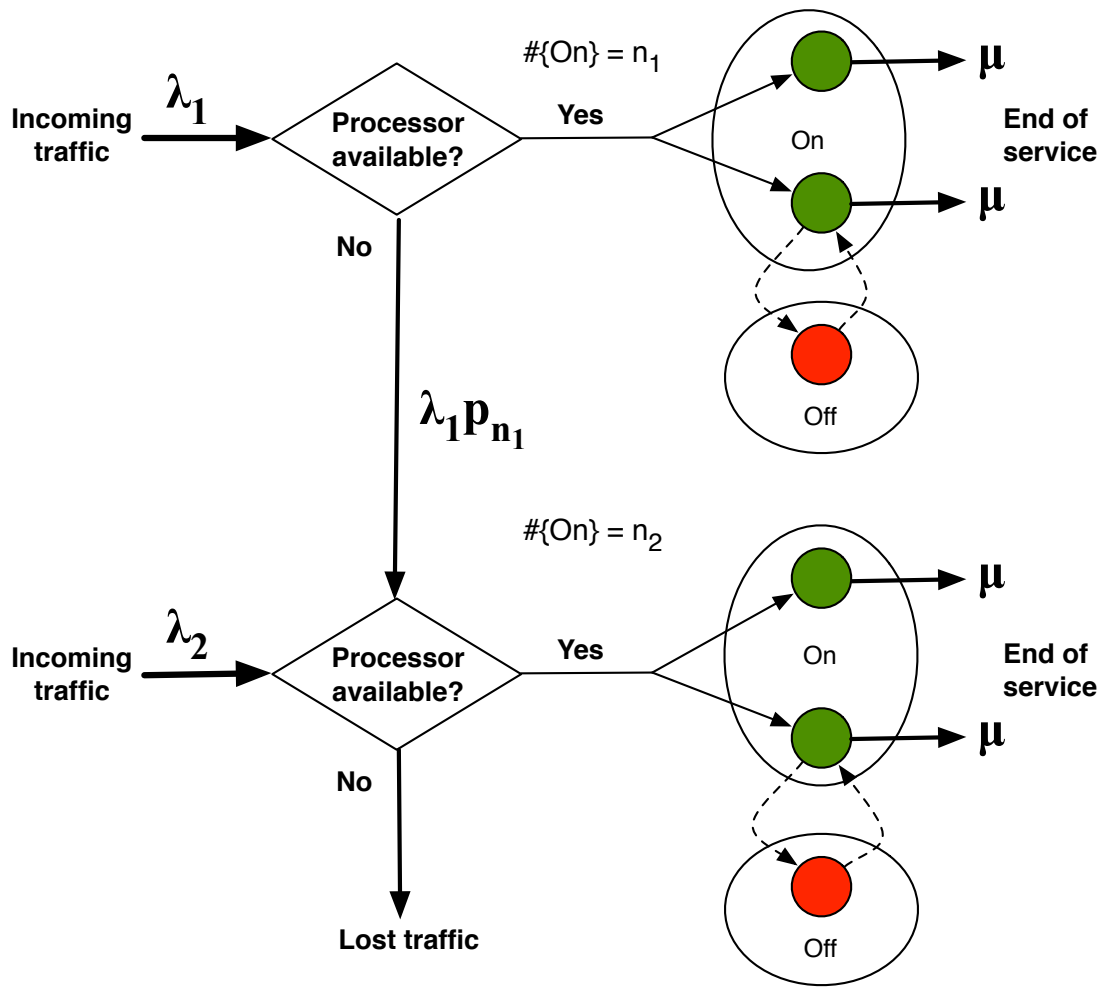

Fig. 1. System model for cloud providers serving two classes of customers. Arriving 'premium' jobs which do not find any available server in pool 1 are offered to the overflow subsystem.

*1) Isolated subsystems:* If the two subsystems are in isolation of each other, they can be treated as two independent $M/GI/n_i/n_i$ queues (the '$M$' stands for Markovian arrivals), which has independent and identically distributed (i.i.d.) service times with a general distribution (the '$GI$') and independent of the arrival process, $n_i$ servers, and no extra waiting spaces (*e.g.*, if all servers are busy, further jobs are lost), augmented with the economic parameters introduced in SectionII-A. Since the Erlang-B model is insensitive to the distribution of job sizes, we do not need to worry about the distribution of job lengths. This model ignores the time-dependence sometimes found in job arrival processes. However, this time-dependence often tends to be not too important over short time intervals.

Under the Erlang loss model, the number of jobs inside the system can be modeled as a Birth-and-Death process with a finite state space, $\{0, 1, \ldots, n_i\}$. Steady state always exists, and the probability $p_{n_i}$ to be in state $n_i$ (*i.e.*, all servers of type $i$ are busy), is given by the Erlang-B formula (see [10] for more details)

$$p_{n_i} = B(n_i, \rho_i) = \frac{\rho_i^{n_i}}{n_i!} p_{0_i}, \quad (4)$$

where $p_{0_i}$ is the probability that subsystem $i$ is empty.

If the arrival process is not Poisson, then the insensitivity property is lost, and the appropriate queueing model becomes $G/GI/n_i/n_i$, for which there is no exact solution. However we can use the approximation described in [11].

Having defined the stationary distribution of the number of jobs present in queue $i$ we can now compute the throughput and the average penalty paid by the provider to premium customers. The average number of type $i$ jobs entering the system (and completing service) per unit time, $T_i$, is

$$T_i = \lambda(1 - p_{n_i}), \quad (5)$$

with $(1 - p_{n_i})$ being the probability that an incoming job finds an idle server. Similarly, the penalty paid to type 'premium' customers, $D$ can be estimated as

$$D = dL_1, \quad (6)$$

where $L_1$ is the rate at which traffic gets blocked at queue 1, *i.e.* $L_1 = p_{n_1}\lambda_1$.

*2) Overflow subsystem:* The revenue earned by the provider per unit time is still given by the sum of the revenues generated by the two subsystems, see Equation (1). However under this model the values of $D$, $T_i$ and $P_i$ differ from those of the first model. In particular, the goal is to increase the system's profitability by lowering either $D$ or $n$ (or both).

Assume that $n_i$ servers have been allocated to queue $i$, while jobs enter subsystem $i$ according to an independent Poisson process with rate $\lambda_i$. As before, operating servers accept one job at a time. However now we require service times to be exponentially distributed with mean $1/\mu$. The second queue is subject to two different traffic streams, *i.e.*, all the jobs of type 2 and the 'premium' jobs which can not be served in the first queue. Hence, even if the Markovian assumptions are met, we do not know the distribution of the traffic offered to the second queue. However a satisfactory approximation is to characterize the $i$th traffic stream by its mean value, $\rho_i$, and its variance, $Var[\rho_i]$. Hence, we make an approximation by considering equivalent two distributions having the same mean and variance.

For the sake of simplicity, assume the existence of a third queue, that is used to run all the type $i$ jobs blocked at queue $i$, $i = 1, 2$. Also, assume that queue 2 has no servers associated to it (*i.e.*, all 'basic' traffic overflows to queue 3), while the $n_2$ servers are allocated to queue 3. The total offered traffic to queue 3 is $\rho_3 = \omega_1 + \omega_2$, where $\omega_i$ is the traffic blocked at queue $i$. We assume that the two traffic streams are independent. Hence we can compute the variance of the traffic offered to queue 3 as the sum of the variance of the two traffic streams, that is

$$Var[\rho_3] = Var[\omega_1] + Var[\omega_2]. \quad (7)$$

If all type 1 servers are busy, then 'premium' jobs are offered to the overflow subsystem with intensity $\omega_1$, otherwise all 'premium' jobs are served by type 1 servers. On the other hand, type 2 jobs are always offered to queue 3, as queue 2 has no servers associated to it (the $n_2$ servers are allocated to queue 3). The average traffic of type $i$ being offered to queue 3 is

$$\omega_i = \rho_i p_{n_i}, \quad (8)$$

while its variance can be computed using the Riordan formula [12]:

$$Var[\omega_i] = \omega_i \left(1 - \omega_i + \frac{\rho_i}{n_i + 1 - \rho_i + \omega_i}\right). \qquad (9)$$

Having computed the mean and variance of the overflow traffic, our next step is to compute the traffic lost in the overflow subsystem when $n_2$ servers are allocated to it and when the offered load has mean $\rho_3$ and variance $Var[\rho_3]$.

Stochastic modeling of overflow loss queues is usually expressed in terms of a multidimensional Markov process. However, unlike the Erlang-B model, the state distribution of these queues does not admit a product-form solution, while numerically solving the balance equations is generally not feasible as the state-space is usually too large. Hence, one has to rely on approximations in order to estimate the blocking probability in overflow loss queues. In this paper we use the approximation proposed by Hayward [13], [14], which tries to describe non Poisson traffic by means of equivalent Poisson traffic, and then apply the usual Erlang-B formula for estimating the blocking probability. Both served and overflow traffic have properties which differ from the the system where the two queues are isolated; however it can be classified according to its *peakedness* $Z$ (a measure of the traffic variability), which is defined as the ratio between the variance and the mean of the number of busy servers in infinite servers queue (see [11] for more details). The peakedness of a Poisson process is 1, while the peakedness of overflow traffic is always greater than 1. When $Z \neq 1$ Hayward approximates the blocking probability as the blocking probability of an Erlang-B system with $n/Z$ servers subject to a load of $\rho/Z$. By applying this transformation the mean and variance of the occupied capacity are the same as those of the original system, while the peakedness becomes equal to 1. When the traffic is Poisson, *i.e.*, when $Z = 1$, this approximation is exact:

$$B(n, \rho, Z) = B\left(\frac{n}{Z}, \frac{\rho}{Z}\right). \qquad (10)$$

While the average load per server does not change, when $Z > 1$ the system becomes smaller, thus incresing the blocking probability. Also, when $Z > 1$ we have to deal with a non-integral number of servers in Equation (10). Therefore we employ the incomplete Gamma function to evaluate the Erlang-B formula.

Having estimated the blocking probability, we obtain the traffic blocked in the overflow subsystem as

$$L = \rho_3 B\left(\frac{n_2}{Z_3}, \frac{\rho_3}{Z_3}\right). \qquad (11)$$

Since our ultimate objective is to determine the amount of jobs that are served, we have to estimate the blocking coefficients for both classes. The two traffic classes do not have the same mean and variance; hence they will not experience the same blocking probabilities in queue 3. In order to estimate the individual blocking probabilities we exploit the observation that the blocking probability for class $i$ is *approximately* proportional to the peakedness of the overflow traffic of type $i$. After some algebraic manipulations we obtain

$$L_i = L\frac{Var[\omega_i]}{Var[\rho_3]}. \qquad (12)$$

Having found the amount of traffic blocked for the two classes, we are now in the position of computing the throughputs, $T_i$, and power consumptions, $P_i$. The average number of jobs of type $i$ entering (and leaving) the system per unit time is

$$T_i = \lambda_i - L_i\mu, \qquad (13)$$

By using the linear power consumption introduced in [3] we can estimate the average power used to run type $i$ jobs as

$$P_i = n_i e_1 + \left\lceil \frac{T_i}{\mu} \right\rceil (e_2 - e_1), \qquad (14)$$

where $T_i/\mu$ represents the average number of type $i$ servers that are not idling, *e.g.*, the average number of type 1 servers running 'premium' jobs and the average number of type 2 servers running either type 1 or type 2 jobs.

**N.B.** The throughput of type 1 jobs (and hence the amount of electricity necessary to execute them) can exceed the maximum nominal throughput. In other words, it is possible to observe $T_1 > n_1\mu$ when some 'premium' jobs are routed to the overflow queue. Similarly, when we allow overflow, $T_2$ is usually lower than the maximum nominal throughput for type 2 jobs, as some type 2 servers are used to run 'premium' jobs.

## III. Policies

In this section we present various allocation policies which allow determining the number of servers required to maximize the provider's profits.

### A. Optimal Server Allocation

Consider an allocation decision epoch, that is, an instant when the allocation decision has to be taken. The state of the system is defined by the two pairs $(n_i, \rho_i)$, $i = 1, 2$. For a given set of demand and economic parameters (the former are estimated from the statistics collected during the observation window), this policy finds a server allocation vector $(n_1, n_2)$[2] which 'maximizes' the total average profit $R = R_1 + R_2$, see Equation (1). We have put quotation marks around the word 'maximizes' because the decisions taken might not be optimal if the exponential assumptions are violated and the queuing model with overflow is used. This allocation policy can be applied anyway but is, in general, a heuristic.

One way of achieving this is to try all possible server allocation vectors and, for each of them, estimated the expected revenue, and choose the best. The number $g$, of different ways that $S$ servers may be allocated between 3 different pools is equal to the number of ways that the integer $S$ can be partitioned into a sum of 3 components. This is equivalent to the number of ways that $S$ indistinguishable balls may be

[2] $n_{off}$ does not need to be acknowledged explicitly, as $n_{off} = S - (n_1 + n_2)$.

allocated into 3 distinguishable boxes [15]. That number is given by:

$$g = \begin{pmatrix} S+2 \\ 2 \end{pmatrix} \approx S^2. \qquad (15)$$

That number grows very quickly with $S$, and the exhaustive search becomes too expensive to the performed on-line, as its computational complexity is in the order of $O(g)$. Hence, we propose the following fast algorithm of the 'hill climbing' variety for maximizing the revenue $R$. A fast search algorithm suggested by the above observations works as follows:

1) Start with some allocation $(n'_1, n'_2, n'_{off})$, *e.g.* by setting $n'_1 = \lceil \lambda_1/\mu \rceil$ and $n'_2 = \lceil \lambda_2/\mu \rceil$, and estimate $R$.
2) Try the four switches where a server is moved from one of the other pools to pool $i$, and the two switches where a server is moved from pool $i$ to one of the other two pools. In each case, evaluate expression (1) with the new vector and choose the best change (*i.e.*, move to the *neighbor* with the highest expected revenue); call that value $newR$.
3) If $newR > R$, set $R = newR$ and $(n'_1, n'_2, n'_{off})$ to the corresponding allocation, and repeat step 2; otherwise go to step 4.
4) Carry out the server allocation $(n'_1, n'_2, n'_{off})$.

It should be noted that hill climbing finds a local optimum, but does not guarantee to converge to the best solution.

If the system is treated as two isolated queues the revenue function is strictly monotonic, as $R$ is a concave function with respect to its arguments $(n_1, n_2)$. Intuitively, the economic benefits of allocating more servers to pools 1 and 2 become less and less significant as $n_1$ and $n_2$ increase. On the other hand the burden for removing servers from pools 1 and 2 gets bigger and bigger as $n_1$ and $n_2$ decrease. Such behavior is an indication of concavity. Therefore this algorithm finds the allocation set which maximizes the total expected revenue.

If we allow overflow, however, the revenue function is discontinuous and non monotonic for some parameters. This is not surprising, as intuitively one should run many servers in order to minimize the probability of paying a penalty, while at the same time he/she should run just a few servers in order to minimize the electricity consumption. Hence, in order to find the optimal solution we propose to evaluate the algorithm for three different initial parameters:

1) $n'_1 = n'_2 = 0$: the server farm does not consume anything, while the paid penalty is $d\lambda_1$.
2) $n'_1 = S$, $n'_2 = 0$: unless the system is highly loaded the provider is rather unlikely to pay any penalty. On the other hand, the expenditure due to electricity consumption is very high.
3) $n'_1 = \lceil \lambda_1/\mu \rceil$, $n'_2 = \lceil \lambda_2/\mu \rceil$: allocate the servers in proportion to the load. This is a reasonable trade-off between the probability to pay penalties and the electricity consumption.

We have run several numerical experiments, and found that by using this heuristic the algorithm always converge to the global maxima.

**N.B** Since powering servers on/off requires time and affects components' reliability, one might easily embed the algorithm we have introduced in [5] into the above policy.

### *B. Heuristic Policies*

The 'Optimal' policy requires requires the evaluation of Equation (1) for several values of $n_1$ and $n_2$. In real settings, especially in large scale deployments, decision making is often done via heuristics, principles and rules of thumb. Hence, it may desirable to have simpler heuristics which, even though not optimal, are easy to implement and allow decisions to be taken faster while still producing good results .

*1) Penalty Capping Heuristic:* Instead of considering the income and expenditure for the next epoch, one might wish to definite an upper bound on the penalties to pay. If that is the case, the following heuristic can be employed:

- From the collected statistics, estimate $\rho_1$ and $\rho_2$. Use the Erlang-B formula, see Equation (4), in order to find the minimum number of servers, $n'_1$, necessary to ensure that the overflow probability of type 1 jobs is less than $\tau$, *i.e.*, $p_{n_1} < \tau$. Since 'premium' jobs which are blocked are offered to the overflow queue, we are putting an upper bound on the number of type 1 jobs that will be lost, *i.e.*, we are minimizing the penalty to pay to blocked customers, $D$.
- We are now left with a load of $(\rho_2 + \omega_1)$ offered to the second queue. Hence, we maximize the expected revenue by solving Equation (3) for different values of $n_2$. Since $R_2$ is a monotonic function when the two queues are treated in isolation of each other , we can use a binary search algorithm in the interval $\{0, \ldots, (S - n_1)\}$ for finding the best $n_2$. Call that value $n'_2$.
- Carry out the allocation $(n'_1, n'_2, (S - n'_1 - n'_2))$.

This algorithm will be referred to as 'Penalty Capping' heuristic.

*2) Percentile Heuristic:* The 'Penalty Capping' heuristic is not as computationally demanding as the 'Optimal' policy. However it does dot directly address the issue of the uncertainty with respect to the arrival rates $\lambda_1$ and $\lambda_2$.

Predictive heuristics based on algorithms such as double exponential smoothing (Winter's method) use historical data in order to try to predict the future workload. However, even after the training period they can not predict the future workload with absolute precision. Hence, in order to deal with the uncertainty with respect to the future load we propose the following heuristic, which performs a slight over-provision aiming at reducing the number of lost jobs (in the following we omit the indexes because this policy deals with one queue at a time, ignoring the overflow):

- From the statistics collected at epoch $k$, estimate the offered load $\rho_k$ using double exponential smoothing. For epoch $(k+1)$ allocate enough servers in order to deal with an error of the forecasting tool, *i.e.*

$$n_{k+1} = \lceil \rho_k + \Delta^x_\lambda \rho_k \rceil. \qquad (16)$$

Now, the problem reduces to finding a value of $\Delta_\lambda^x$ capable of reducing job loss without wasting too much energy. More formally, we are trying to ensure that the probability of underestimating the future arrival rate is lesser than a certain probability:

$$Pr(\lambda < (1 + \Delta_\lambda^x)\hat{\lambda}) = x, \quad (17)$$

where $\hat{\lambda}$ is the predicted arrival rate for the next epoch, and $\Delta_\lambda^x$ is the $x$-th percentile of the relative error, which can be easily estimated from the cumulative distribution function (CDF) of the relative error obtained from the historical data.

Even though several techniques allow to compute that estimate for certain forecasting algorithms, our aim here is to derive simple rules which can be easily deployed, rather than providing 'exact' results which have not much practical interest. Hence, for each epoch $k$, we compute $Var(\lambda)$ by recording the relative difference between the predicted and the actual arrival rates, *i.e.*, $\Delta_k = (\hat{\lambda}_k - \lambda_k)/\lambda_k$.

In [6] we have analyzed the relative forecasting error for Winter's method applied to the Wikipedia workload of November 2009 and found that the 95-*th* percentile produced by double exponential smoothing is 0.11. It is perhaps worth noting that the variance of the relative error does not significantly change over the time. In other words, by setting $\Delta_\lambda^{0.95} = 0.11$ in Equation (16) we are 95% sure that the system will not be overloaded. We have experimented with different percentile values and we have found that the 95-*th* percentile guarantees the best trade-off between energy consumption and lost jobs. This policy will be referred to as 'Percentile' heuristic.

In scenarios where the workload is dynamic and hard to predict, a logical extension of the 'Optimal' policy might be to invoke the 'Optimal' allocation algorithm with the parameters suggested by the 'Percentile' heuristic.

## IV. Experimental Evaluation

In this section we present a number of experiments which were carried out in order to understand how the proposed policies affect various metrics. We assume the server farm has a Power Usage Effectiveness (PUE), the main metric used to evaluate the efficiency of data centers, of 1.7. That value is computed as the ratio between the total facility power and the IT equipment power. The other settings are summarized in Table I. The last hypothesis we make is that jobs are not completely CPU bound. Instead, when a job executes, it requires 70% of the CPU time, on average. Since the power consumption is a linear function of the CPU usage we can easily determine the power consumption of busy servers as $e_1 + 0.7 \times (e_2 - e_1) = 76.15$ Wh.

In the first set of experiments we describe, for both the model with overflow and that with no overflow, how the allocation and the revenue change when the load is fixed but the amount of penalty to pay to 'premium' customers for failed transactions varies between 0 and 5 \$. As depicted in Figure 2(a), when the two subsystems are isolated, the allocation policy simply increases the number of running servers as the monetary penalty increases. On the other hand, if we allow overflow the best strategy is to have maximum flexibility by allocating all the servers to queue 2 (this maximizes the total throughput), but only if the penalty is 'low'. If the penalty grows above a certain threshold, see Figure 2(b), then the flexibility does not pay off anymore. Therefore the best strategy is to start differentiating the traffic. The higher the penalty, the higher becomes $n_1$. The total number of server increases as well, but not to the same extent. In other words, the number of servers allocated to the overflow queue decreases when the penalty increases.

| | | |
|---|---|---|
| $S$ | 1,000 | Number of servers |
| $r$ | 0.1 \$ KWh | Electricity cost |
| $c_1$ | 0.03 \$/hour | Charge (type 1 jobs) |
| $d$ | 0.2 \$ | Penalty (type 1 jobs) |
| $c_2$ | 0.085 \$/hour | Charge (type 2 jobs) |
| $1/\mu$ | 2.5 hours | Average service time |
| $e_1$ | 59 Wh | Power consumption (idle) |
| $e_2$ | 83.5 Wh | Power consumption (busy) |
| PUE | 1.7 | Power usage effectiveness |

TABLE I
Settings.

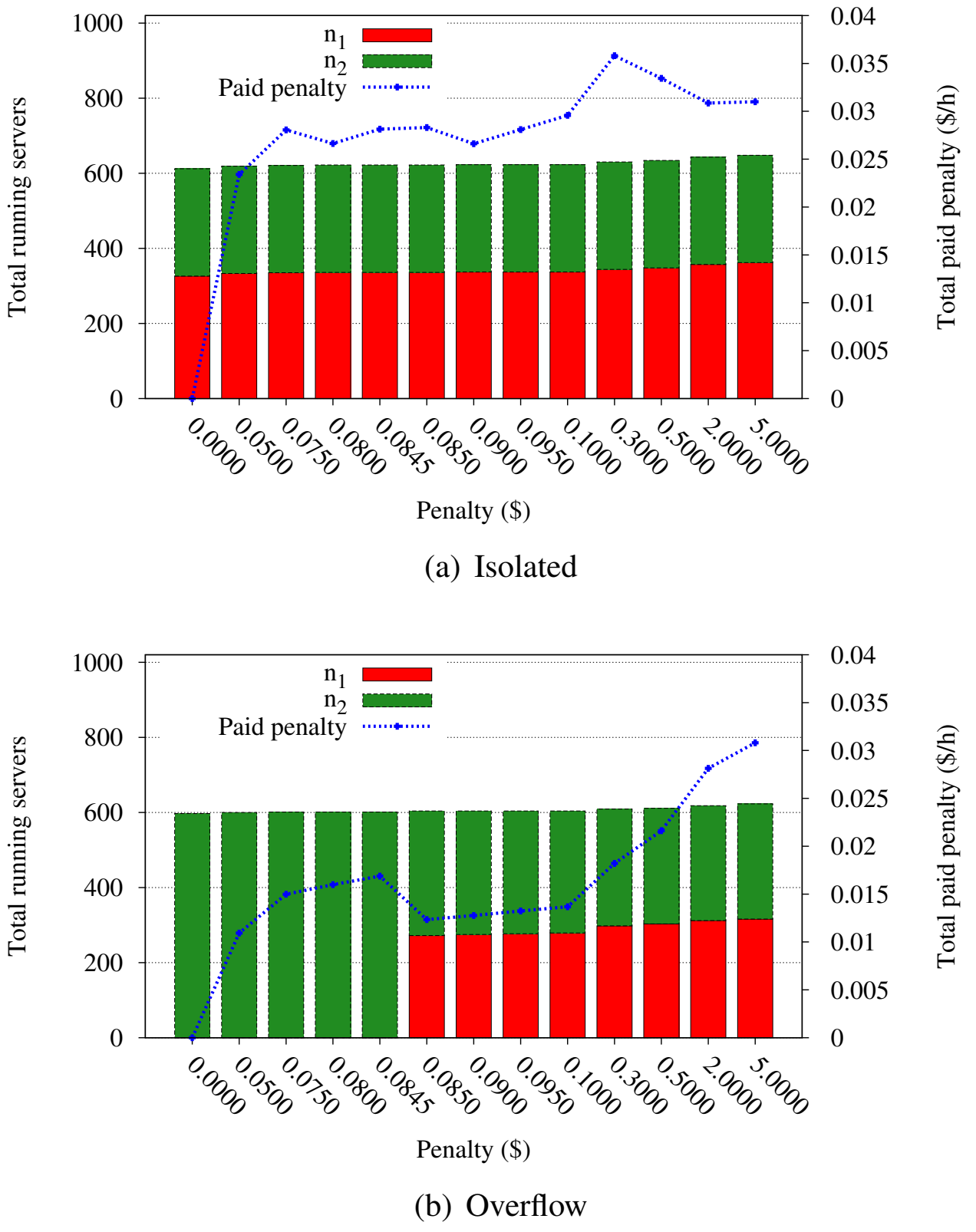

Fig. 2. Optimal allocation for (a) isolated and (b) overflow models for different values of $d$. $\rho_1 = 300$, $\rho_2 = 250$, other settings as in Table I.

In Figure 3 we compare the maximum achievable revenue for the 'Overflow' and 'Isolated' models. For comparison reasons we also display the revenues achieved by the 'Always On' policy, *i.e.*, a policy which always runs $n_2 = S$ servers. In other words, all 'premium' traffic overflows to the second queue, where it competes with type 2 jobs for the available resources. Given that the total offered load is relatively low (the system is 55% loaded) the probability of paying a penalty is negligible. However running too many servers negatively impacts the performance obtained by the 'Always On' policy, as high electricity cost erode revenues. On the other hand, by carefully choosing the number of servers to run, the 'Isolated' model performs almost as good as that allowing overflow.

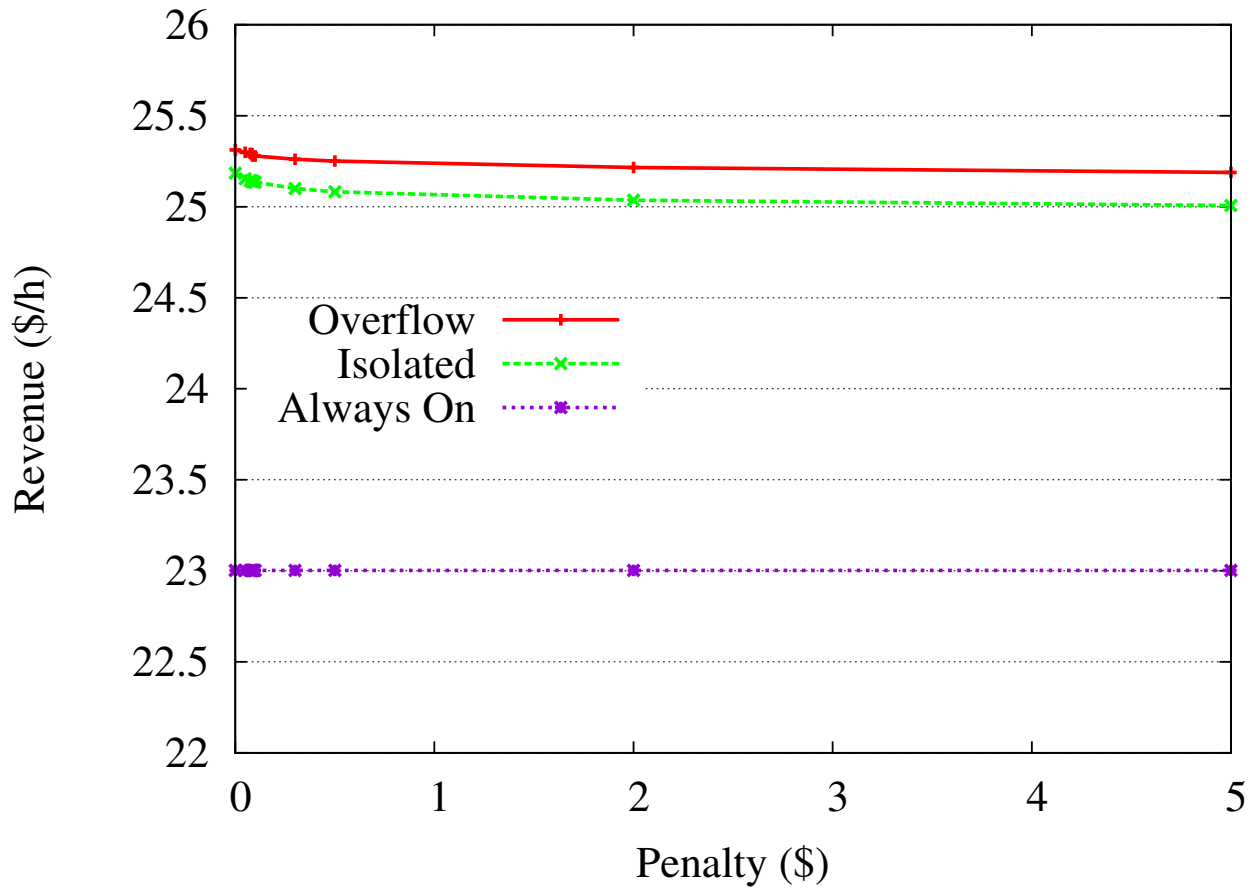

Fig. 3. Achieved revenue for different values of $d$. $S = 1000$, $\rho_1 = 300$, $\rho_2 = 250$, other settings as in Table I. The 'Always On' policy runs all the servers and uses a common pool, while the dynamic policies use the best allocation, *i.e.*, those displayed in Figure IV.

Next, we experiment with different electricity costs, *i.e.*, $r = \{0, \ldots, 4\}$ \$/kWh. As once might expect, when the cost for electricity is negligible the best strategy is to over-provision the system in order to reduce the probability of failing to run a transaction of type 1 due to the lack of resources, see Figure IV. As the electricity cost increases more and more servers are turned off, thus increasing the number of failed transactions and ultimately the amount of penalty to pay to 'premium' customers, see Figure 5. Finally, beyond a certain threshold it becomes more sense from a financial point of view to pay penalty to all 'premium' customers rather than running servers. Hence, for $r > 1$ kWh it is better for the provider to turn all the servers off.

Next, we evaluate the two models we have described and, for the model allowing overflow, the performance of the policies we have introduced in Section III by departing from the assumption that the load is known and stationary. Since we are not aware of any publicly available data describing user demand of Cloud resource we have extrapolated it from the available Wikipedia traces. In the following experiment we use a scaled version of the Wikipedia traffic of November 2009 [3]. The traces exhibit a general trend with different patterns (*e.g.*, weekly and daily) as well as unexpected spikes, which make the future load rather difficult to predict. Incoming jobs enter the system according to two Poisson processes with rate $\lambda_1$

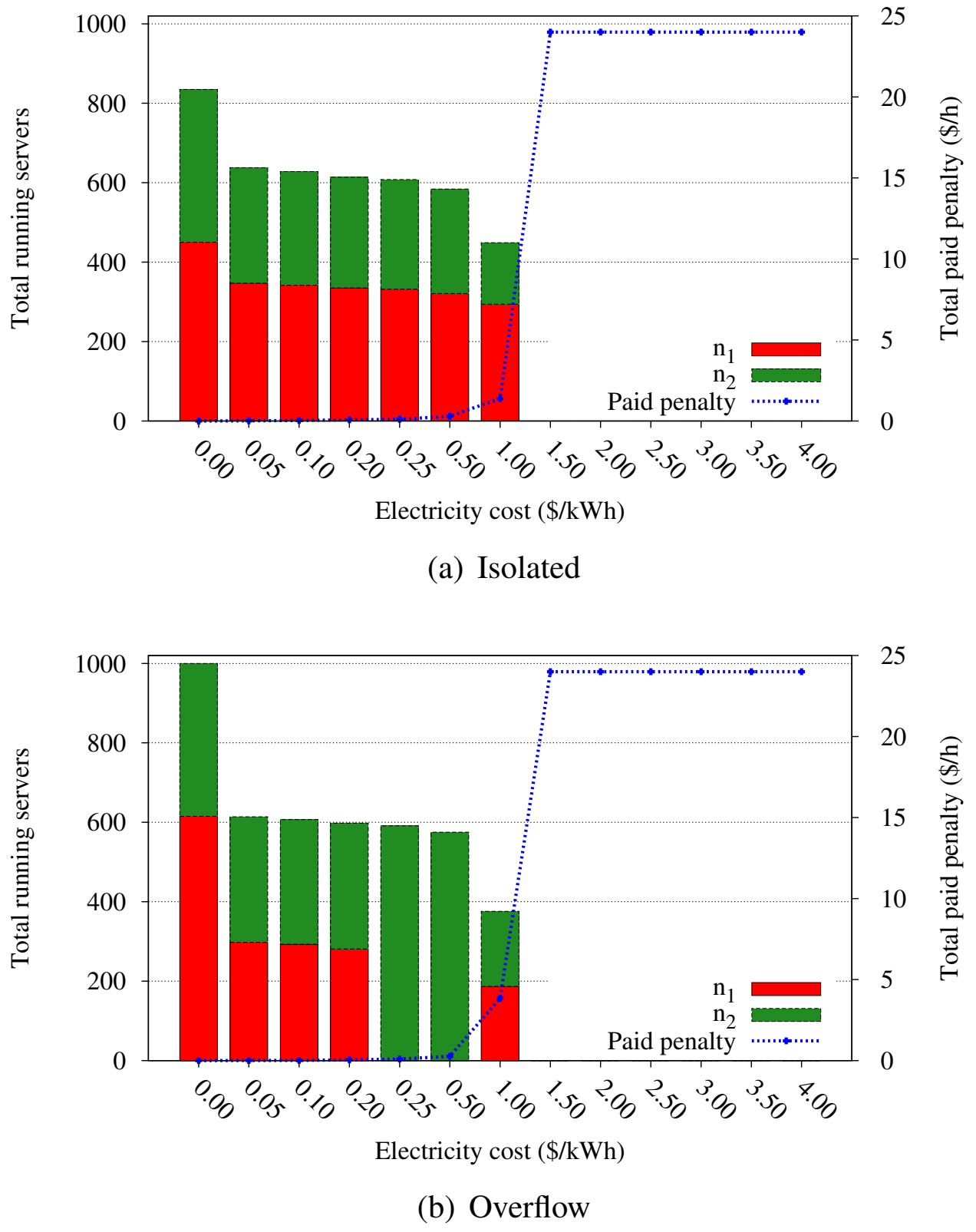

Fig. 4. Optimal allocation for (a) isolated and (b) overflow models and (c) achieved revenue for different values of $r$. $\rho_1 = 300$, $\rho_2 = 250$, other settings as in Table I.

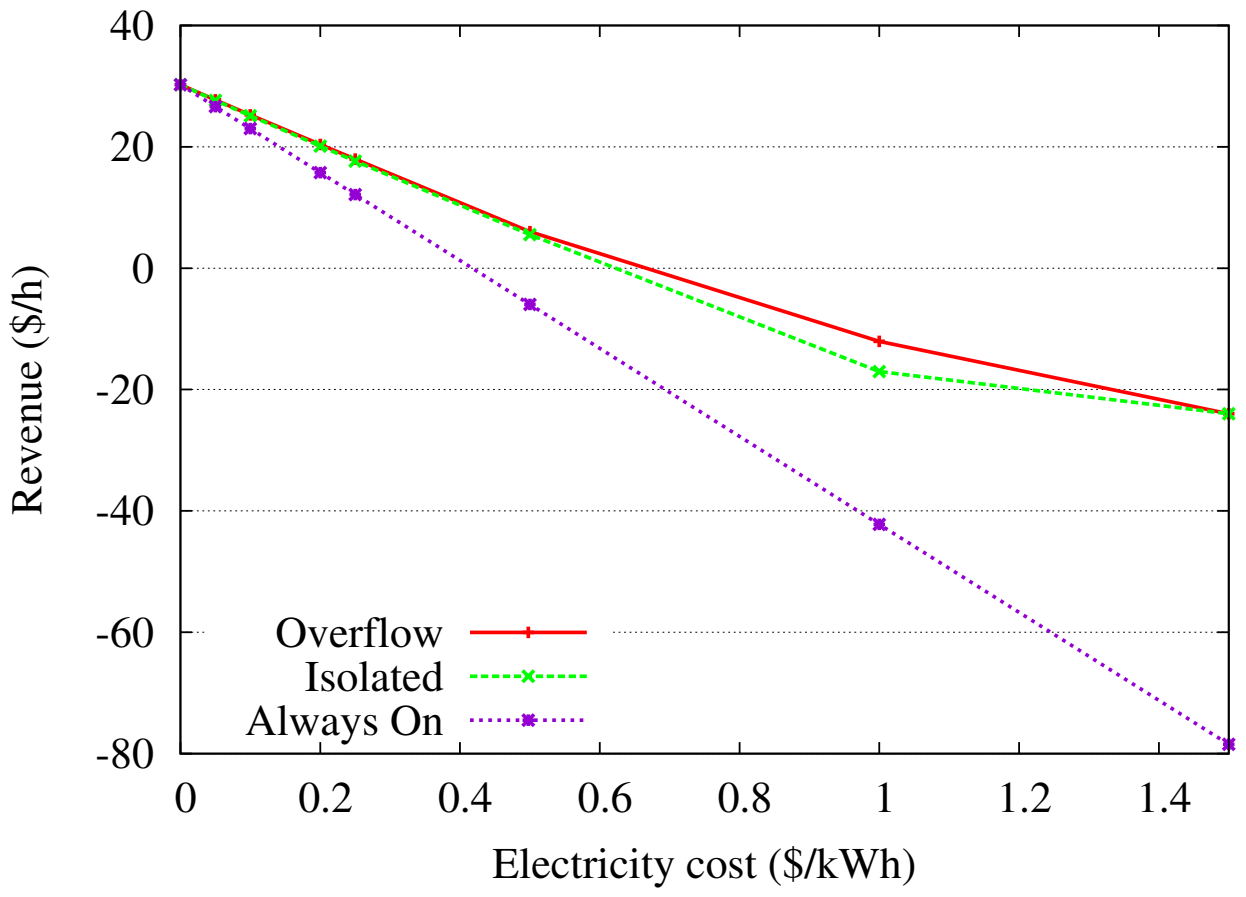

Fig. 5. Achieved revenue for different values of $r$. $\rho_1 = 300$, $\rho_2 = 250$, other settings as in Table I. The 'Always On' policy runs all the servers and uses a common pool, while the dynamic policies use the best allocation, *i.e.*, those displayed in Figure IV. For $r \geq 1.5$ \$/kWh the best strategy is to turn all servers off and to pay penalties to all 'premium' customers.

and $\lambda_2$, which change every hour, while the allocation policy is invoked every two hours.

In order to make the make the model more realistic, we also take indirect costs into account. These include the cost for capital as well as the amortization of the equipment such as servers, power generators or transformers, and account for twice the cost of consumed electricity. As one can see in Figure 6, the two heuristics perform almost as good as the 'Optimal' policy. Probably the most important property of the two policies is that, even though they perform a slight over-provision which entails a higher power consumption, they are capable of significantly reducing the number of lost jobs. The 'Penalty Capping' heuristic does so because we have required the blocking probability for 'premium' jobs to be less than 0.001%, while the 'Percentile' heuristic achieves similar results by smartly under-estimating the parameters suggested by the Winter's method. Please note, however, that a 50% improvements in the number of lost jobs is not a large number in absolute terms and constitutes less than 1% of the total traffic (the 'Optimal' policy loses about 2% of both 'premium' and 'basic' jobs).

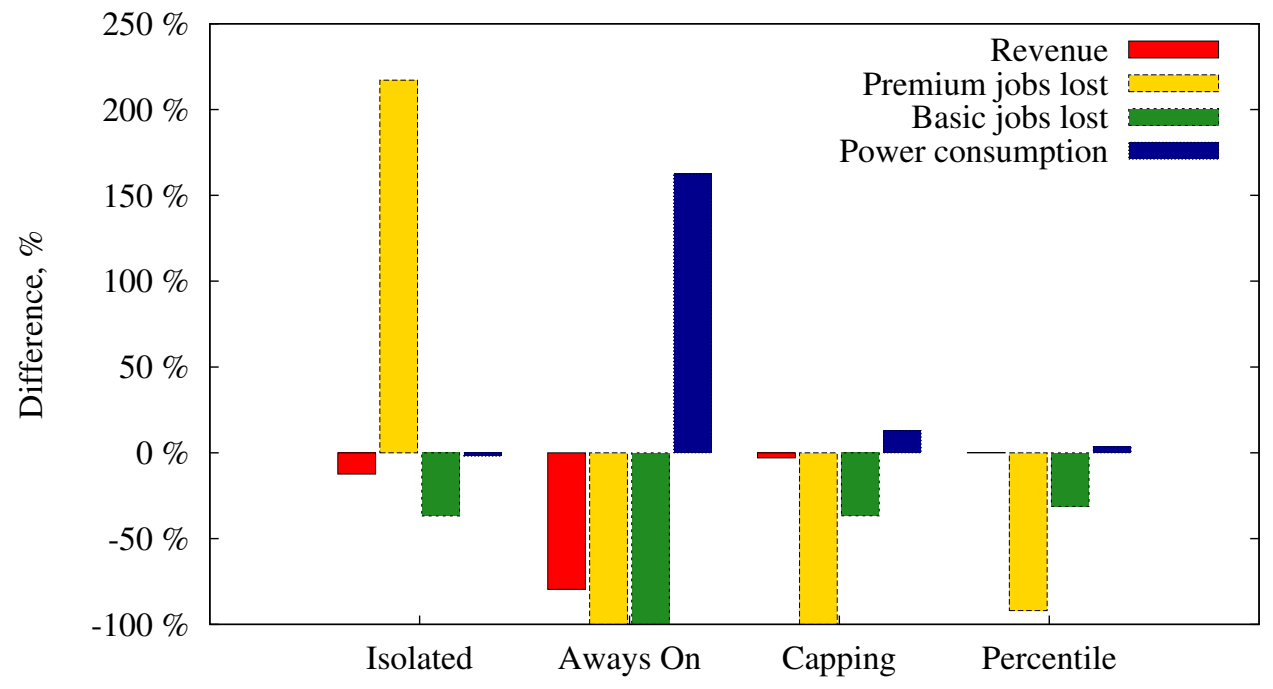

Fig. 6. Performance of the heuristics compared to that obtained by the 'Optimal' policy over a one month period. The experiment uses a scaled version of the Wikipedia traffic of November 2009. Settings as in Table I.

From the figure we observe that the 'Always-On' policy achieves a decrease in job losses by running all the servers. This negatively impacts the achieved revenue, as they are eroded by the high energy consumption. On the other hand, the 'Isolated' model is a good approximation in terms of achieved revenues. However it loses too many 'premium' jobs.

## V. Conclusions

This paper proposed multiple models for determining the revenue-maximizing number of servers to be allocated for two classes of customers in a server farm: premium customers who make upfront reservations and are entitled to demand service availability, and basic ones who are served on a best-effort basis. Two types of models were considered: one where the two pools of allocated servers ('premium' and 'basic') are kept 'isolated', and another where the instances allocated for basic customers can be used for premium customers if needed, but not vice-versa ('overflow' model).

The experimental evaluation has shown that the number of running servers and the number of servers allocated to each pool have a significant effect on the revenue earned by the provider. The optimal allocation across pools is highly dependent on the penalty. With low penalties, the allocation where all servers are placed in a single pool (and used to serve both 'premium' and 'basic' customers) is preferable. With higher penalties, more servers need to be allocated to serve only 'premium' customers. Also, the optimal allocation is dependent on the energy cost. The experiments also show that the proposed models, particularly the overflow models, adapt well to different traffic conditions. According to the experimental results, both the 'Penalty Capping' and the 'Percentile' heuristics are good candidates for practical implementation in scenarios where the load changes unexpectedly over the time.

Possible directions for future research include taking into account the trade-offs between the number of running servers, the frequency of the CPUs, and the maximum achievable performance, the power consumed by the networking equipment (*i.e.*, switches), as well as fault tolerance issues.

## Acknowledgements

The authors would like to thank Dmytro Dyachuk for the useful comments and discussions, and Mick Flanagan for providing an implementation of the Erlang-B formula allowing us to deal with non-integer resources[3]. This work was partly funded by the EU Regional Development Funds under EUREKA Project 4989 (SITIO) and EU Cost Action IC0804.

[3]http://www.ee.ucl.ac.uk/~mflanaga/java/